\documentclass[11pt]{article}
\usepackage{epsfig,amsfonts}
\usepackage[fleqn]{amsmath}
\usepackage{amsthm,amssymb}
\usepackage{graphicx}
\usepackage{hhline}
\usepackage{cite}

\def\be{\begin{equation}}
\def\ee{\end{equation}}
\def\bdm{\begin{displaymath}}
\def\edm{\end{displaymath}}
\def\bea{\begin{eqnarray}}
\def\eea{\end{eqnarray}}

\newcommand{\rd}{\mbox{d}}
\newcommand{\ri}{\mbox{i}}
\newcommand{\re}{\mbox{e}}

\begin{document}

\begin{titlepage}
\begin{flushright}
RU-NHETC-2003-20\\
\end{flushright}

\vspace{1.5cm}

\begin{center}
\begin{LARGE}
{\bf Integrable Circular Brane Model and Coulomb
Charging at Large Conduction}

\end{LARGE}

\vspace{1.3cm}

\begin{large}

{\bf Sergei L. Lukyanov} {\bf and}
{\bf Alexander B. Zamolodchikov}
\end{large}

\vspace{.5cm}

{NHETC, Department of Physics and Astronomy\\
     Rutgers University\\
     Piscataway, NJ 08855-0849, USA\\

\vspace{.2cm}

and

\vspace{.2cm}
L.D. Landau Institute for Theoretical Physics\\
  Chernogolovka, 142432, Russia}
\vspace{1.5cm}

\centerline{\bf Abstract} \vspace{.8cm}
\parbox{11cm}{
We study a model of $2D$ QFT with boundary interaction, in which
two-component massless Bose field is constrained to a circle at the
boundary. We argue that this model is integrable at two values of the
topological angle, $\theta =0$ and $\theta=\pi$. 
For $\theta=0$ we propose
exact partition function in terms of solutions of ordinary
linear differential equation. 
The circular brane model is equivalent to
the model of quantum Brownian dynamics 
commonly used in describing the
Coulomb charging in quantum dots,
in the limit of small  dimensionless
resistance $g_0$ of the tunneling contact. Our proposal
translates to partition function of this model at integer charge.
}
\end{center}

\begin{flushleft}
\rule{4.1 in}{.007 in}\\
{June 2003}
\end{flushleft}
\vfil

\end{titlepage}
\newpage

\section{ The model}

In this note we address a model of $2D$ Euclidean field theory
with boundary interaction which we call the circular brane model.
The model contains two-component Bose field ${\bf X}(z,{\bar z}) =
\big({\rm X}^{1}(z,{\bar z}), {\rm X}^{2}(z,{\bar z})\big)$
defined on a disk of
a radius $R$ with usual complex coordinates $(z,{\bar z})$,
$|z| \leq R$. Its bulk dynamics is that of a free massless
field, as described by the action
\bea\label{action}
{\cal A}_{\rm disk} = {1\over  \pi}\,\int_{|z|\leq R} \rd^2 z
\ \partial_{z}{\bf X}\cdot \partial_{\bar z}{\bf X}\
\eea
($\rd^2 z = \rd{\rm x} \rd{\rm y}$), but it obeys nonlinear
boundary condition
\bea\label{circle}
{\bf X}^2_B ={\textstyle{ 1\over  g_0}}\ ,
\eea
where
${\bf X}_B$ stands for the values of ${\bf X}$ at
the boundary $|z|=R$, and  $g_0$ is a coupling constant. Due to the
nonlinear
boundary condition, the theory needs renormalization. It has to be
equipped with the ultraviolet (UV) cutoff, 
and consistent removal of the
UV divergences requires that the bare coupling constant $g_0$ be
given a dependence of the cutoff momentum $\Lambda$, according to
the Renormalization Group (RG) flow equation
\bea\label{flow}
\Lambda\, {{\rd g_0}\over{\rd\Lambda}} = - 2 \, g_{0}^2 - 4
\,g_{0}^3 + \cdots\ .
\eea
The leading two terms of the beta 
function written down in\ \eqref{flow}\
were
computed in\ \cite{Kosterlitz}\ and\ \cite{Zwerger},
and indeed agree with the more general
calculations in\ \cite{Leigh}\ and\ \cite{Rychkov}. 
Eq.\eqref{flow}\
shows that the theory
is asymptotically
free -- at short 
distances the effective size of the circle \eqref{circle}\
becomes
large. It is believed that the 
theory develops a physical energy scale
$E^{*}$, and at the distances $\gg 1/E^{*}$ from 
the boundary the
effect of the boundary is that of the fixed boundary condition ${\bf
X}_{B} =0$ (the boundary\ \eqref{circle}  ``flows'' to the Dirichlet
boundary). According to the equation\ \eqref{flow}
\bea\label{scale}
E^*\sim \Lambda
\ g_{0}^{-1}\,\re^{-{1\over{2g_0}}}\ .
\eea

As usual, the effect of the boundary can be described in
terms of the boundary state\ \cite{nappi}. The boundary state $|\, B\,
\rangle$ is
a special vector in
the space of states of radial quantization (in our case the space of
states of two-component free massless Bose field on a circle of the
circumference $2\pi R$).
This state incorporates all information
about the boundary conditions; any correlation function can be
written in terms of the Euclidean-time Heisenberg operators sandwiched
between $|\, B\, \rangle $ 
and the radial-quantization vacuum $|\, {\bf
0}\, \rangle$.
In particular, the overlap $\langle\,  {\bf 0}\, |\, B\, \rangle$
coincides with the disk partition function, up to the factor $R^{c/6}$. 
In our case $|\,
{\bf 0}\, \rangle$ is the Fock 
vacuum of the free massless field ${\bf
X}(z,{\bar z})$ with the zero-mode momentum ${\bf P}={\bf 0}$. More
generally, we will be interested in 
the overlap amplitude $\langle\,  {\bf
P}\,|
\, B\,  \rangle$, where $|\, {\bf P}\,  
\rangle$ is the Fock vacuum with
the zero-mode momentum ${\bf P}$; of course, this amplitude relates
to the (unnormalized)
one point correlation function of the exponential field
inserted at the center of the disk,
\bea\label{sbdgfr}
\langle\, {\bf P}\, |\, B\, \rangle =R^{-c/6+{\bf P}^2/2}\ \, \langle\, 
\re^{\ri{\bf
P}\cdot{\bf X}}(0,0) \,   \rangle_{\rm disk}\ .
\eea
Here $c=2$ is the central charge of the bulk theory. 
It will be convenient to transform the spherical brane model to
the geometry of semi-infinite cylinder, using the standard
exponential map:
\bea
z/R = \re^{u/R}\ ,\quad
\quad {\bar z}/R  = \re^{{\bar u}/R}\ ,
\eea
where $u = \sigma + \ri\tau, \ {\bar u} =
\sigma - \ri\tau$, with $\sigma$
ranging from $-\infty$ to $0$, and $\tau$ being a periodic
coordinate with the period $2\pi R$. In what follows we will write
(with some abuse of notations) ${\bf X}(\tau,\sigma)$ for the field
${\bf X}$ in \eqref{action}
expressed in terms of these cylindrical
coordinates. Then, introducing the
shifted field variable
\bea
{\bf Y}(\tau,\sigma) = 
{\bf X}(\tau,\sigma) +{\ri \sigma \over R}\  {\bf
P}\, \ ,
\eea
one can relate the the 
amplitude $\langle\,  {\bf P}\, |\,   B\, \rangle$
to
the partition function\footnote{The factor 2 in\ \eqref{partit}\ is added
partly to 
eliminate the spurious boundary degeneracy of the auxiliary 
Neumann boundary at $\sigma
= -\ell$ in \eqref{acyl}, and partly to take into account the infrared 
boundary entropy 
of the real boundary at $\sigma=0$; with this normalization the $R\to\infty$
behavior of $Z(R,H)$ is $\sim \re^{-2\pi R\,{\cal E}}$, with the coefficient
1.}:
\bea\label{partit}
Z(R,H) = \lim_{\ell\to\infty}\bigg\{ 2\,\re^{{\ell\over 6R}}\, 
\int\,{\cal D}{\bf Y} \ \re^{-{\cal A}_{\rm cyl}[{\bf
Y}]}\, \bigg\}\ ,
\eea
where the integration field variable ${\bf Y}(\tau,\sigma)$ is still
subject to the circular brane constraint
\bea\label{bc}
{\bf Y}^2_B  = {\textstyle{ 1\over  g_0}}\ ,
\eea
and
\bea\label{acyl}
{\cal A}_{\rm cyl} [{\bf Y}] =&& {1\over
4\pi}\,\int_{-\ell}^0\rd\sigma\oint \rd\tau
\ \big(\partial_{\sigma}{\bf Y}\cdot \partial_{\sigma}{\bf Y}
+\partial_{\tau}{\bf Y}\cdot
\partial_{\tau}{\bf Y}\big)\\ && -2\, \oint \rd\tau\, {\bf H}\cdot
{\bf Y}_B\, .\nonumber
\eea
Here and below $\oint d\tau$ stands for $\int_{0}^{2\pi R} d\tau$.
The ``external field'' ${\bf H}$ relates to ${\bf P}$ in
\eqref{sbdgfr} as
\bea
{\bf H} ={  { {\rm i}\, {\bf P}\over {4\pi R}}}\ ,
\eea
and the argument $H$ in the r.h.s. of\ \eqref{partit}\ is
$\sqrt{{\bf
H}^2}$. Thus
\bea\label{shsdyy}
\langle\, {\bf P}\, |\,  B\, \rangle =
{\textstyle{1\over \sqrt 2}}\ Z\big(\, R\, ,\,  {\textstyle{\sqrt{-
{\bf P}^2}\over 4\pi
R}}\, \big)\ ,
\eea
where the factor ${\textstyle{1\over \sqrt{2}}}$ is 
the boundary degeneracy\ \cite{affleckl}\ of the infrared 
(Dirichlet) fixed point. We note that by the nature of the
boundary condition\ \eqref{bc}\ the partition function $Z(R,H)$ is
expected to be
an entire function of $H$, and its analytic continuation to pure
imaginary $H$ is unambiguous. In the form\ \eqref{acyl}\ the model has
interpretation as 1+1 QFT on a 
half-line, at the temperature $T = {\textstyle{1\over 2\pi
R}}$. At $R\to\infty$ the partition function\ \eqref{partit}\ develops
standard
linear asymptotic
\bea\label{sbdyt}
\log Z(R,H) \to - 2\pi R\ {\cal E}(H) \quad {\rm as}\quad R\to\infty\ ,
\eea
where ${\cal E}(H)$ is the boundary energy. 
This quantity depends on $H$
and
$g_0$, the latter dependence being through the energy scale $E^*$,
Eq.\eqref{scale}. We found it 
convenient to fix overall normalization of
$E^{*}$
relating it to the zero temperature susceptibility,
\bea\label{succept}
{1\over{E^{*}}} = - {1\over 2}\ {{\partial^2 {\cal E}}\over{\partial H
^2}}\Big|_{H=0}\ .
\eea

The circular brane model shows many similarities with the $O(3)$ sigma
model (or ${\bf n}$-field), the asymptotic freedom\ \eqref{flow}\ being
just
one of them\ \cite{APolyakov}.
Another similarity is apparent when one observes that the
field configurations ${\bf Y}(\tau,\sigma)$ in\ \eqref{partit}\  can be
separated
into the topological classes characterized by the integer-valued
winding number
\bea\label{widing}
w = {g_0\over{2\pi}}\,\oint\rd\tau\  { {\bf Y}_B\wedge
\partial_{\tau}{\bf Y}_B}\ ,
\eea
just like the sigma-model field configurations ${\bf n}$ fall into
their topological classes characterized by the mapping degree 
${\Bbb S}^2
\to {\Bbb S}^2$.
Moreover, in the sigma-model
the instanton configurations
minimizing the action in each topological sector are found explicitly
in terms of rational 
functions\ \cite{Belavin}; very similar expressions
exist for
the instanton configurations in the circular brane model\
\cite{Korshunov,Nazarov}.
The calculations of the
instanton determinants\ \cite{Frolov}\ also exhibit much similarity\
\cite{Wang,Feigelman,Larkin}.
As usual,
existence of the integer-valued topological charge allows one to
introduce another parameter, the topological angle $\theta$,
\bea\label{parttheta}
Z_{\theta} = \sum_{w\in{\Bbb Z}}\,Z^{(w)}\,e^{i\theta w}\ ,
\eea
where the partition functions $Z^{(w)}$ are the functional
integrals\ \eqref{partit}\ taken over the field
configurations in the topological class $w$ only.
The full partition function $Z$, 
as defined by Eq.\eqref{partit}, involves
all
topological sectors with equal weights, i.e. $Z=Z_{\theta=0}$. It is
this $\theta=0$ partition function which we had in mind in defining the
boundary energy ${\cal E}(H)$ and the energy scale $E^{*}$ through\
\eqref{sbdyt},\eqref{succept}.

One of the reasons we mention this analogy here is because the $O(3)$
sigma  model is known to be integrable at two special values of the
topological angle, $\theta = 0$ and $\theta = \pi$
\cite{Polyak,ZZ,ZF,ZZA}. Therefore it is
natural to expect that the circular brane model is also integrable at
these two special values of $\theta$
\footnote{It is worth mentioning here that the
case $\theta=\pi$ is of special interest. In this case the circular
brane model flows to a nontrivial conformal boundary
condition which is believed to be related
to the infrared fixed point of two-channel spin-${1\over 2}$
Kondo model in its interacting sector\ \cite{Matveev}.}.
We argue in  Appendix that
for these values of $\theta$ the model admits a number of nontrivial
higher-spin local integrals of motion. In fact, our argument there
goes for the whole class of the ``$O(N)$ spherical brane models'',
with $N$-component field ${\bf X}$, subject to the spherical brane
boundary condition \eqref{circle}. Moreover, in these models it is
possible to
describe the whole set of the local 
integrals of motion ${\Bbb I}_{2k-1}$,
$k=1,2,3\ldots$ in some details.

The aim of this note is to present a proposal for the exact form of
the partition function \eqref{partit} at the integrable point
$\theta=0$ in terms of solutions of certain
ordinary differential equation. We will present this in the next
section, but let us
first make some remark and introduce suitable notations. As was
mentioned above, renormalization trades the bare coupling constant
$g_0$ for the RG
invariant scale\ \eqref{scale}, and the partition
function actually depends on the dimensionless combination $R E^{*}$.
In fact, in the circular brane model (as well as in the $O(3)$ sigma
model) the last statement is not entirely true. As was observed in
\cite{Wang}, the functional integral\ \eqref{partit}
has a specific nonperturbative divergence due to the small-size
instantons, which cannot be absorbed into the renormalization of the
coupling constant $g_0$. Due to this effect,
the partition function\ \eqref{partit}\ is expected to have the form
\bea\label{sclepart}Z_{\theta} = e^{L \kappa\, \cos \theta}\
 Z^*_\theta(\kappa, h)\ ,
\eea
where
\bea\label{consta}
L =A\ \log\big(B(g_0)\, {\Lambda/ E^{*}}\big)\ ,\
\eea
and
\bea\label{juyy}
\kappa=2\pi R E^{*}\, ,\ \ \ \ \ \ \ \ \ \  h=2\pi RH\ .
\eea
Here $A$ is a constant, which
is universal (actually $A=2$), but $B(g_0)$ is not -- it depends on
the details of UV cutoff. The factor $ Z^*_\theta$
depends  on the RG invariant parameters $\kappa,\, h$ and $\theta$ only.
Similar nonperturbative small-instanton divergence is well known to
be present in the $O(3)$ sigma model \cite{Frolov},
where it is blamed, for instance, for the violation of the normal
scaling of the topological susceptibility \cite{Lusher}.
It is important to note that the small
instanton divergence does not 
affect the $h$-dependence of the free energy
(interaction of small instantons with the external field ${\bf H}$ is
negligible), therefore the energy scale $E^{*}$ defined through
\eqref{sbdyt} and
\eqref{succept} enjoys the normal RG behavior\ \eqref{scale}.
In what follows we discuss the
universal
factor $Z^* (\kappa,h)$ in the full partition function at $\theta =0$,
\bea\label{sbdgt}
Z(R,H)= \re^{ \kappa L}\ Z^* (\kappa,h)\,,
\eea
which carries all interesting dependence 
on $H$ and $R$. Since any factor
of the form $\re^{const\ \kappa}$ 
can be absorbed into the redefinition of
$B(g_0)$ in \eqref{consta}, we fix the ambiguity by assuming that
$Z^* (\kappa,0) \to 1$ as
$\kappa\to\infty$.
With this convention, the
boundary
energy can be written as
\bea\label{shdyoiu}
{\cal E}(H) = LE^* + {\cal E}^* (H)\,,
\eea
where ${\cal E}^*(0) = 0$.

\section{The partition function}

It was discovered some years ago in Ref.\cite{toteo} that in the
case of the boundary flow in  the minimal CFT with
integrable boundary perturbation\ \cite{blz}, the overlap analogous to
\eqref{sbdgfr}\ can be related exactly to the eigenvalue problem
of certain ordinary differential operator.
Despite the fact that this
relation was proven\ \cite{blzz}, and similar relations were found in
other
integrable models of CFT with non-conformal boundary interactions\
\cite{baza,Tsvelik},
its deeper reason  remains mystery to us. Nonetheless, it looks
natural to assume that it might be a general phenomenon, and try to
identify associated differential operator for the circular
brane model. Below we simply present our
proposals for such differential operator. The motivations came from
detailed studies of somewhat
more general, but still integrable ``brane'' model (we call it the
``paperclip brane model''); it generalizes the circular model in a way
similar to the ``sausage'' deformation of the $O(3)$ sigma
model\ \cite{sausage}. We will report these studies elsewhere\ \cite{luvz}.

Consider the following ordinary differential equation
\bea\label{equat}
\Big\{\, -\partial_{v}^2 + \kappa^2\, \exp\big({\re^{v}}\big)
+ h^2\,\re^{v}\, \Big
\}\, \Psi(v)
=0\ .
\eea
Let $\psi_{-}(v)$ be the solution which decays at large negative $v$,
\bea\label{kdju}
\psi_{-}(v) \to \re^{\kappa v} \quad \quad
 {\rm as}\quad \quad
 v \to -\infty\ .
\eea
Also, let $\psi_{+}(v)$ be the solution decaying at large positive
$v$,
\bea\label{smdjhhy}
\psi_{+}(v) \to \exp\bigg\{-{ \re^{v}\over 4} + \kappa
\ {\rm Ei}\Big({\re^{v}\over
   2 }\Big)\, \bigg\} \ \  \quad
{\rm as}\ \  \quad
 v \to +\infty\ ,
\eea
where ${\rm Ei}(x)$ is the integral exponent function, 
${\rm Ei}(x) =
{\rm P.V.}\int_{-x}^{\infty}\, {\textstyle
{{\rm d} t\over t}}\, \re^{-t}$.
The asymptotics
\eqref{kdju}\ and\ \eqref{smdjhhy}\ specify
the solutions $\psi_{-}(v)$ and $\psi_{+}(v)$ uniquely,
including their normalizations. Then
\bea\label{resi}
{Z}^* (\kappa, h) =\sqrt{\pi\over \kappa}\ \
{{(2\, \re^{\gamma_E-2}\, \kappa^2)^{\kappa}}
\over{\Gamma(1+2\kappa)}}\ \
  W\big[\, \psi_{+}\, ,\, \psi_{-}\, \big]\ ,
\eea
where $W$ is the Wronskian $\psi_{+}\, \partial_{v}\psi_{-}-
\psi_{-}\,\partial_{v}\psi_{+}$ 
and $\gamma_E=0.577216\ldots$ is Euler's
constant.
Thus determination of ${ Z}^*$
reduces to the problem of the ordinary differential equation.

\section{Low temperature expansion}

At large $R$ (low temperatures) the equation\ \eqref{equat}\ can be
studied using
WKB expansion. The leading WKB approximation yields explicit expression
for the boundary energy\ \eqref{shdyoiu},
\bea\label{fsdsre}
{\cal E}^* (H) =-E^*\  \int_0^{\infty} {\rd t\over t}\ 
\Big(\sqrt{\re^t+ t
\, (H/E^*)^2}
-\re^{t/2}
\, \Big)\, .
\eea
Note that the integral admits the  expansion in powers of $H^2$,
\bea\label{qwwydy}
{\cal E}^* (H) =E^*\ 
 \sum_{k=1}^{\infty}\,C_k \ \big({\textstyle{ H\over E^*}}\big )^{2k}\,,
\eea
with
$$C_k =
{(-1)^k\over 2 \sqrt{\pi}}\ \
{\Gamma(k-{1\over 2})\over k}\ \ \big(k-{\textstyle{{1\over 2}}}\big)^{-k}
\ .$$
This expansion should be understood in terms of the expansion
\eqref{Bexp}, with
the powers $H^{2k}$ representing the highest order
terms
$h^{2k}$ of the polynomials $I_{2k-1}$, the vacuum eigenvalues of the
local IM ${\Bbb I}_{2k-1}$. Note that the above expression for the
coefficients
$C_k$ is consistent with\ \eqref{snydyy}.
The next order of the WKB expansion provides leading correction to the
linear asymptotic\ \eqref{sbdyt},
\bea\label{lowT}
\log Z^* =- 2\pi R\,{\cal E}^*  - {1\over 2\pi R}\ {\cal E}^*_2
+O(R^{-3})\ ,
\eea
where, by direct calculation,
\bea
{\cal E}^*_2 =
{1\over 48 H}\ \, {\partial\over
\partial H} \Big(\, H\ {\partial{\cal E}^*\over \partial H}
\, \Big)\ .
\eea
Again, expanding\ \eqref{lowT}\ in powers of $H$ we have
\bea
\log Z^*\simeq -\sum_{k=1}^{\infty}\,C_k \ 
(2\pi E^*)^{1-2k}\ \ I_{2k-1}\ ,
\eea
with
\bea\label{bbst}
I_{2k-1} = {1\over{R^{2k-1}}}\,\Big(\, h^{2k} + {k^2\over
12}\ \, h^{2k-2}+\ldots\, \Big)\,,
\eea
in perfect agreement with\ Eqs.\eqref{Bexp},\eqref{sbbsgt}.
Further terms in \ \eqref{bbst} can be obtained by computing yet
higher orders of the WKB expansion. This way one finds
\bea
&&I_1= {1\over R}\,\Big(\, h^2 + {1\over 12}\, \Big)\,,\nonumber\\
&&I_3
= {1\over R^3}\,\Big(h^4 + {1\over 3 }\ h^2 +
{1\over 40}\Big)\,,\\
&&I_5= {1\over R^5}\,\Big(\, h^6 + {3\over 4}\ h^4+
{19\over 100}\ h^2
+{71\over 5040}\, \Big)\, ,\nonumber
\eea
again in agreement with \ \eqref{imsv}.

\section{High temperature expansion}

It is also possible to develop 
short distance (high temperature) expansion
of the partition function\ \eqref{resi}.
For small $\kappa$ the solutions
$\psi_{-}(v)$
and $\psi_{+}(v)$ can be evaluated using appropriate versions of
perturbation
theory. To make the resulting formulae compact we shall describe here
these
expansions in the case $h = 0$ only.
The solution $\psi_{-}(v)$ can be written as
\bea\label{sajsuy}
\psi_{-}(v) = \re^{\kappa v}\ \Big\{ 1 +  \sum_{n=1}^{\infty} \,
a_n (\kappa) \ \re^{nv}\,  \Big\}\,,
\eea
where the coefficients $a_n (\kappa )=O(\kappa^2)$
and admit systematic expansions in
powers
of $\kappa$. The $\kappa \to 0$ expansion 
of the solution $\psi_{+}(v)$ is
more subtle. Instead of powers of $\kappa$, it expands in powers of the
``running coupling constant'' $g=g(\kappa)$ defined by the equation
\bea
\kappa = g^{-1}\ \re^{-{1\over {2g}}}\ .
\eea
After change of variables
\bea
v = g \,x - \log (g)
\eea
the equation\ \eqref{equat}\  becomes
\bea\label{perturbat}
\Big\{ - \partial_{x}^2 + \re^x + \delta U (x)\, \Big\}\Psi =0\, ,
\eea
where
\bea
\delta U(x) =
\exp\Big({{\re^{gx}-1}\over g}\Big) - \re^x\,.
\eea
For $|x| \sim 1$ the term $\delta U(x)$ is $O(g)$; in this
domain solutions of \ \eqref{perturbat} admit systematic expansion in
powers of $g$.
Expansion of $\psi_{+}$ is obtained by iterations of\
\eqref{perturbat},
starting with
$\psi_+^{(0)} = C(g)\ K_0 \big(2\, \re^{x/2}\big)$, where $K_0$ is
the Macdonald function and  the normalization constant $C(g)$ must be
adjusted to match the asymptotic form\ \eqref{smdjhhy}. It turns out that
both expansions, of $\psi_{-}$ in the powers of $\kappa$, and of
$\psi_{+}$ in the powers of $g$, 
have common domain of validity at $v \sim
-\log (g)$, and can be used there for evaluation of the
Wronskian in\ \eqref{resi}.
As the result, the following form of the small $\kappa$ expansion of the
partition function\ \eqref{sbdgt}\ emerges
\bea\label{instant}
Z(R,0)\simeq \re^{2\kappa\, \log(2\pi B(g_0) R\Lambda)}\ \
{{g^{\kappa}}\over{\sqrt{g}}}\ \sum_{n=0}^{\infty}\,\kappa^n\
z_n
(g)\,,
\eea
where the coefficients $z_n (g)$ are power series in $g$. The term $n=0$
in\ \eqref{instant}\ should be interpreted as the perturbative
contribution to the
partition function, the term $n=1$ corresponds to the one-instanton
contribution, and the 
higher nonperturbative terms $n=2,3\ldots$ presumably
describe the contributions of multiple instanton and anti-instanton
configurations. Explicit computation along the lines described above
yields
\bea
&&z_0 (g) = 1 - (1+\gamma_E )\ g + O(g^2)\,, \label{zets}\\
&&z_1 (g) =\log(2)-2+3\gamma_E-
2\, \gamma_E\, g-2\, (2+\gamma_E)\, g^2+O(g^3)\ . \nonumber
\eea

\section{Dissipative quantum rotator}

The circular brane  model has useful interpretation in terms of
Brownian dynamics of quantum rotator. It was noticed a while ago in
Ref.\cite{Calan}\ that the
free massless bulk dynamics\ \eqref{acyl}\ is equivalent to the
Caldeira-Leggett model of quantum thermostat\ \cite{Legett}.
Upon fixing the boundary
values ${\bf Y}_B(\tau)$
and integrating out the  bulk part of
the field ${\bf Y}(\tau,\sigma)$,
Eq.\eqref{partit}\
reduces
to
\bea\label{partdissip}
Z(R,H) = \int\,{\cal D}\eta\
e^{-{\cal A}_{\rm diss}[\,\eta\, ]}\ ,
\eea
with
\bea\label{disact} {\cal A}_{\rm diss}[\,\eta\,]& =&
-{2H\over \sqrt{g_0}}\,\oint\rd\tau\,\cos(\eta) +\\ &&\ \
{1\over{8\pi^2R^2 g_0 }}\ \oint \rd\tau\,\oint
\rd\tau'\ {{\sin^2\big(\, {{\eta(\tau)-\eta(\tau')}\over 2}\, \big)
}\over\sin^2 ({\tau-\tau'\over 2R})
}\ ,\nonumber
\eea
where $\eta(\tau)$ is the angular 
field defined through ${\bf H}\cdot{\bf
  Y}_B (\tau) = H/\sqrt{g_0}\ \cos \eta(\tau)$. The model
similar\footnote{The model of \cite{Eckern} has the potential
term $\cos(2\eta)$ instead of $\cos(\eta)$ in \eqref{disact}.}
to\ \eqref{disact}\
was
introduced in\ \cite{Eckern}\ (see e.g. \cite{Zaikin}  for a review)
as an effective
field theory describing tunneling of quasiparticles between
superconductors. More recently it was subject of much interest in
studying of the phenomenon of Coulomb blockade in quantum dots\
\cite{Averin}, in the
regime where the dimensionless  resistance  of the tunneling contact
$g_0$
is small\
\cite{Falci,Wang,Zwerger,Feigelman,Larkin}.
In this context, the effective action in fact contains another term,
\bea\label{cutoff}{\cal A}_{\rm CB} = {\cal A}_{\rm diss} +
  {1\over{4E_C}}\,\oint \rd\tau\,\eta^2_{\tau}\ ,
\eea
where $E_C$ has the meaning of the  charging energy of the
dot in the absence
of the tunneling,
$E_C = e^2/2C$. Of course, $\tau$ is the Matsubara time, and $2\pi R =
1/T$, the inverse temperature. When $T\ll E_C$ this term just provides
explicit UV cutoff, with the cutoff energy
\bea\label{cutoffaa}
\Lambda =  E_C/g_0\ .
\eea
With the cutoff procedure thus specified, the ``small-instanton''
factor $\re^{\kappa L}$ in \ \eqref{sbdgt} becomes unambiguous. By
comparing \eqref{instant},\eqref{zets} with direct one-instanton
calculation in
\ \eqref{cutoff}\ \cite{Wang,Feigelman,Larkin}\ one 
finds for small $g_0$
\bea\label{lgnot}
L={\textstyle{1\over g_0}}+5\, \log(g_0)
+O(1)\ .
\eea
Eqs.\eqref{sbdgt},\eqref{resi} then provide the partition function
of the model \ \eqref{cutoff} at $\theta = 0$, in the limit of small
$g_0$. One only has to relate the energy scale $E^*$ in \ \eqref{juyy}
to the parameters
$g_0$ and $E_C$ in the above action\ \eqref{cutoff}.
When $E_C /g_0 \gg H \gg E^*$ and $g_0$ is small, the boundary energy
${\cal E}(H)$\ \eqref{shdyoiu}\ can 
be computed directly from \eqref{cutoff},
using the standard
perturbation theory in $g_0$. By developing two-loop order in this
perturbative expansion
and comparing it with the $H/E^* \gg 1$ behavior of the
integral in\ \eqref{fsdsre},
we found
\bea\label{nsshyhy}
E^* = {{E_C}\over{2\pi^2\,g_{0}^2}} \ {\rm e}^{-{1\over{2 g_0}}}\,
\Big\{\,  1 - {{3\pi^2}\over 4}\,g_0 + O(g_{0}^2 )\, \Big\}\,.
\eea

Finally, let us make a remark on the topological susceptibility in the
model\ \eqref{cutoff}.
As was already mentioned, the topological susceptibility
\bea\label{asiiu}
E_C^*=-{\pi\over R}\ {\partial^2\over
\partial\theta^2}\, \log (Z_\theta ) \Big|_{\theta=H=0}
\eea
in the circular brane model
suffers
from the small-instanton divergence, which leads to the non-universal
factor
$\re^{\kappa L\,\cos\theta}$ in\ \eqref{sclepart}.
On the other hand, in applications
to
the Coulomb blockade, the topological susceptibility in
\eqref{cutoff} has direct
physical
meaning -- it coincides
with measurable capacitive
energy  of the quantum dot. Explicit cutoff in the
action\ \eqref{cutoff}
makes
the small-instanton factor unambiguous. According to\ \eqref{sclepart},
at zero
temperature
and zero $H$, $E_{C}^*$ has the form $2\pi^2\, (L + const )\, E^*$,
where the
constant  comes from
the $\theta$-dependence of the second factor in\ \eqref{sclepart}.
Since our proposal is
strictly limited to the case $\theta=0$, it
does not allow for determination of this constant.
However, at small $g_0$
the dominating contribution comes from the term $L E^*$,
and therefore
in view of\ \eqref{lgnot} and \eqref{nsshyhy},
the charging energy at zero temperature is estimated as
\bea\label{Capacitive}
E_{C}^*\, |_{T=0} = {{E_C}
\over{g^2_0}}\ \re^{-{1\over 2g_0}}\
\Big\{\, {1\over{g_0}} + 5\,\log g_0 +
O(1)\, 
\Big\}\ .
\eea
Of course, the leading behavior ($1/g_0$ in the brackets) here is
consistent with the one-instanton analysis
in the Gaussian approximation\ \cite{Wang,Feigelman,Larkin}.
Nontrivial
prediction
which follows from our proposal  is the logarithmic term
in\ \eqref{Capacitive}.

\section{On other spherical brane models}

The circular brane model is a particular case $N=2$ of
the ``$O(N)$ spherical
brane model'', the latter containing
$N$-component field ${\bf X}$, subject
to the same boundary condition\ \eqref{circle},
so that ${\bf X}_B \in {\Bbb S}^{N-1}$\ \cite{Kosterlitz}. As we
argue in Appendix, the spherical brane model is integrable for any
$N \geq 1$. Although the bulk part of this
paper is devoted to the
circular
case $N=2$, here we would like to say few words about other
cases. First,
for
generic $N\neq 2$ there are no topological sectors, 
and the small-instanton
divergence
is not present. For $N\neq 2$ the overlap\ \eqref{sbdgfr}\ $(c=N)$ can be written as
\bea\label{qswsu}
\langle\, {\bf P}\, |\, B\, \rangle=2^{-{N\over 4}}\  \re^{-\kappa C_0}\ Z^* (\kappa, h)\,,
\eea
where $\kappa$ and $h$
are still the dimensionless parameters\ \eqref{juyy},
and $E^*$ is defined  in Eq.\eqref{succept}.
The factor $Z^* (\kappa, h)$ here   
is normalized as in\ \eqref{sbdgt},
i.e.
$Z^* (\kappa,0) \to 1 $ as $\kappa \to \infty$, and
the $N$-dependent constant  $C_0$ is 
the same as in
Eqs.\ \eqref{Bexp},\eqref{frexp}.

For generic $N$ we were not able to find differential equations which
could
possibly generate the overlap\ \eqref{qswsu},
as\ \eqref{equat}\ does for $N=2$.
However,
for two other cases, $N=1$ and $N=3$, representations of the function
$Z^*$
in\ \eqref{qswsu}
in terms of solutions of certain differential equations exist. Here
we would like to describe what these differential equations are.

For $N=1$, it is useful to note that the model is equivalent to the
interacting sector of the one-channel spin-${1\over 2}$
Kondo model. At  $N=1$ the sphere degenerates to two points, which
represent two states of the impurity spin.
Exact equivalence is
established by bosonizing the Kondo fermions 
(see e.g. Ref.\cite{affleck}).
As is well known, the
Kondo model is solvable by means of Bethe 
ansatz technique\ \cite{wtsv,aflo}. Alternative
(and equivalent) representation of the partition function can be given
in terms of the solutions of the differential equation
\cite{Tsvelik}
\bea\label{esdsdquat}
\Big\{\, -\partial_{v}^2 + 2\pi\kappa^2\ v\, {\re^{v}}
- h^2\, \Big
\}\, \Psi(v)
=0\ .
\eea
The function $Z^*$ coincides with associated Stokes multiplier
(see\ \cite{Tsvelik}
for the details).

For $N=3$, consider the differential equation
\bea\label{esdsdquataa}
\Big\{\, -\partial_{v}^2 + \pi\kappa^2\  {\re^{v^2/ 2}}
+ h^2\, \Big
\}\, \Psi(v)
=0\ .
\eea
Let $\psi_{+}$ and $\psi_{-}$ be its solutions uniquely determined by
the asymptotics
\bea\label{en3}
&&\psi_{+} \to \exp\Big\{-{v^2\over 8}-\pi\kappa\ {\rm Erfi}(v/2) 
\Big\}\ \ \  \quad
 {\rm as}\  \quad
 v\to+\infty\, ,\\
&& \psi_{-} \to \exp\Big\{-{v^2\over 8}-\pi\kappa\ {\rm Erfi}(-v/2)
\Big\}\  \quad
  {\rm as}\  \quad
 v\to-\infty\, ,\nonumber
\eea
where\ ${\rm Erfi}(x)$
is the imaginary error function, ${\rm Erfi}(x) =
{\textstyle{2\over \sqrt{\pi}}}\ \int_0^x
\rd t\, \re^{t^2}$. The
suggested relation is
\bea\label{zn3}
Z^* (\kappa, h) = (2\kappa\sqrt{\pi})^{-1}\ 
\, W\big[\, \psi_{+}\, ,\, \psi_{-}\, \big]\,.
\eea
Like in the case $N=2$, the $\kappa\to\infty$ expansion of
\eqref{zn3}\ can be
obtained by WKB analysis of the solutions of
\eqref{en3}. Thus, the leading WKB
approximation yields the boundary energy\ \eqref{frexp},
with the coefficients $C_k$
given exactly by\ \eqref{snydyy} with $N=3$.
Moreover, higher orders of the WKB expansion
reproduce  the eigenvalues\ \eqref{imsv},\eqref{sbbsgt}\ perfectly.

\bigskip
\section*{Acknowledgments}

The authors would like to acknowledge useful conversations
with L.B. Ioffe and  M.V. Feigelman.
SLL is grateful to  A.M. Tsvelik for interesting
discussions.
ABZ thanks Al.B. Zamolodchikov and P.B. Wiegmann for
sharing their insights.

\bigskip

\noindent
The research is supported
in part by DOE grant $\#$DE-FG02-96 ER 40949
SLL also acknowledges a support  from Institute
for Strongly Correlated and Complex Systems at BNL
where the final part of this work was done.

\bigskip
\bigskip
\section{Appendix}

\subsection{The ${ O(N)}$ spherical brane model}

In this appendix we present arguments in favor of integrability of
the circular brane model at $\theta = 0$ and $\theta = \pi$. In
fact, the arguments apply without much modification to more general
model, which contains $N$-component Bose field ${\bf X} =
\big(X^{1}, \cdots , X^{N}\big)$,
with $N \geq 1$, but otherwise is described by the same action
\eqref{action}\ and
the ``spherical brane'' boundary condition \eqref{circle}.
Therefore throughout
this appendix we mostly address this $N$-component spherical brane
model. Presence of arbitrary number of components allows for
a number of checks within the $1/N$ expansion.

\subsection{Integrals of motion}

In the bulk, the spherical brane model is a free Bose field, and
as such it certainly has infinite number of integrals of motion. The
derivative $\partial {\bf X}$, where $\partial \equiv \partial_u$
stands for
the derivative over $u = \sigma + \ri\tau$,
is a holomorphic field in the bulk.
Hence any local polynomial $P(u) = P(\partial{\bf X},
\partial^2 {\bf X}, \cdots)$ of this and higher derivatives with
respect to $u$ gives rise to a local integral of
motion
\bea\label{ip}
{\Bbb I}[P] = \oint{\rd\tau\over 2\pi}\ 
P(\partial{\bf X}, \partial^2 {\bf
X},
\ldots)
\ ,
\eea
where we have assumed the coordinate $\sigma$ along the cylinder to be
the Euclidean time. Here and below
the bulk composite fields entering these polynomials are always
understood in terms of the normal ordering with respect to the
standard Wick pairing $\langle\, {\rm X}^{a}(u,{\bar u})\, {\rm X}^b
(u',{\bar u}')\, \rangle = - \delta^{a b}\,
\log|u-u'|$ corresponding to a bulk free field in an infinite space.
Certainly, there is a ``left-moving''
counterpart to each of the integrals\ \eqref{ip},
\bea\label{ibarp}
{\bar {\Bbb I}}[P] =\oint
{\rd\tau\over 2\pi}\ P({\bar\partial}{\bf X}, {\bar\partial}^2 {\bf
X},\ldots)
\ ,
\eea
where ${\bar\partial} \equiv \partial_{\bar u}$.
The integrals\ \eqref{ip},\eqref{ibarp}\ are operators acting in the
space of states
of free $N$-component massless Bose field quantized on a spatial
circle of the circumference $2\pi R$, i.e.
\bea\label{space}
{\cal H} = \int_{\bf P} \,{\cal F}_{\bf P}\otimes {\bar{\cal F}}_{\bf
P}
\ ,
\eea
where ${\cal F}_{\bf P}$ is the Fock space of the right-moving bosons
with the zero-mode momentum ${\bf P}$.
Large set of integrals \eqref{ip}\ contains many infinite subsets
${\Bbb I}[P_s]$ (corresponding to special sequences of the polynomials
$P_s$)
of mutually commutative operators,
\bea\label{comm}
\big[\, {\Bbb I}[P_s]\,  ,\,  {\Bbb I}[P_{s'}]\, \big] = 0\ .
\eea
The polynomial fields $P_s$ can always be chosen to have definite
spin, and it is convenient to label them accordingly; below we always
assume that $s$ indicates the spin of $P_s (u)$. We will also use
conventional notation ${\Bbb I}_{s-1}$ for the integral ${\Bbb I}[P_s]$.

As is discussed in \cite{gz}, integrability of the field theory\
\eqref{action}\ in
the presence of the boundary requires that the boundary state
satisfies the equations
\bea\label{ib}
(\, {\Bbb I}_s - {\bar {\Bbb I}}_s\,  )\  |\,  B\,  \rangle = 0\ ,
\eea
for all ${\Bbb I}_s$ in one such subset. 
These equations must follow from
special properties of the boundary conditions, which should be such
that the corresponding fields $P_s(u)$ and $P_s ({\bar u})$,
being brought to the boundary $\sigma =0$, satisfy the conditions
\bea\label{deriv}
\big(\, P_s (u) - P_s ({\bar u})\, \big)\big|_{\sigma=0} 
=\ri\,  {\rd\over
\rd\tau}\, \Theta_s
(\tau)\ ,
\eea
where $\Theta_s(\tau)$ are some (renormalized) 
local boundary fields. In
the
case
of the spherical brane boundary condition\ \eqref{circle}, existence of
the first
few polynomials satisfying \eqref{deriv} can be demonstrated by an
argument
similar to that previously used to support integrability of the
$O(N)$ sigma model \cite{Polyak}. The $O(N)$ symmetry of the boundary
condition \eqref{circle} suggests that the polynomials $P_s$ must have
this
symmetry as well. We thus can look for the corresponding integrals
${\Bbb I}_{s-1}$ in the form:
\bea
&&{\Bbb I}_1 = \oint {\rd\tau\over 2\pi}
\,  \partial {\bf X}\cdot \partial {\bf X}\,,\nonumber\\
&&{\Bbb I}_3 =
\oint {\rd\tau\over 2\pi}\,  
\Big[\, \big(\partial{\bf X}\cdot\partial{\bf
X}\big)^2 +
b_3\,\big(\partial^2 {\bf X}\cdot\partial^2 {\bf X}\big)\,
\Big]\,,\label{PTKJ}\\
&&{\Bbb I}_5 = \oint {\rd\tau\over 2\pi}\,
\Big[ \big(\partial {\bf X}\cdot\partial {\bf X}\big)^3 +
b_5\, \big(\partial^2 {\bf X}\cdot \partial^2 {\bf X}\big)
\big(\partial {\bf X}
\cdot\partial {\bf X}\big)\nonumber\\ &&\ \ \ \ \ \ \ +
c_5\,  \big(\partial^2
{\bf X}\cdot\partial{\bf X}\big)^2 + d_5\,  \big(\partial^3 {\bf
X}\cdot\partial^3 {\bf X}\big) \Big]\, ,\nonumber
\eea
and for generic $k$
\bea\label{sjssuyy}
{\Bbb I}_{2k-1} &&= \oint{\rd\tau\over 2\pi}
\Big[\big(\partial {\bf X}\cdot \partial {\bf X}\big)^k + b_{2k-1}
\big(\partial^2 {\bf X}\cdot\partial^2 {\bf X}\big)\, 
\big(\partial {\bf X}
\cdot\partial {\bf X}\big)^{k-2} \nonumber
\\ && +
c_{2k-1}
\big(\partial^2 {\bf X}\cdot\partial {\bf X}\big)^2
\big(\partial {\bf X}
\cdot\partial {\bf X}\big)^{k-3}+
\ldots
 \Big]\, .
\eea
Note that in writing these
expressions we have fixed the normalization of the currents $P_{2k}$:
the term $\big(\partial{\bf X}\cdot\partial{\bf X})^k$, 
having the highest
power of ${\bf X}$,
comes with the coefficient 1. In what follows we always assume this
normalization of the integrals ${\Bbb I}_{2k-1}$.
The terms omitted in \ \eqref{sjssuyy} have powers of
${\bf X}$ equal $2k-4$ or lower.
In general case, the
difference $\big(\, P_s (u) - P_s ({\bar u})\,
\big)\big|_{\sigma=0}$ is a
combination of local boundary fields having appropriate symmetries.
These include the $O(N)$ symmetry and the anti-symmetry with respect to
the reflection $\tau \to -\tau$. Note that for $N=2$ the last
reflection symmetry is violated unless $\theta =0$ and $\theta = \pi$.
Also, only the fields of right
scale dimensions (i.e. equal or below the scale dimension of $P_s$) are
admitted. Elementary counting shows that for $s=2$ the only admissible
field is a total derivative over $\tau$, of the form of the
r.h.s. of \eqref{deriv}. For $s=4$ there is only one admissible field
which
is not a total derivative, and for $s=6$ there are three
admissible non-derivative fields. It follows that $P_2$
always satisfy \eqref{deriv}, and the coefficients   in\ \eqref{PTKJ}
can be
adjusted to ensure that $P_4$ and $P_6$ satisfy\ \eqref{deriv} as well.
The
role of the boundary condition\ \eqref{circle}\ in this counting is in
reducing
the number of independent boundary fields: the boundary conditions
determine the number of redundant boundary fields. As
usual, this kind of argumentation fails for higher spins, since the
number of admissible fields grows too fast.
Nonetheless, in many other
models similar ``low-spin test'' was successful in detecting full
integrability, with no known (to us) exceptions. Therefore we regard the
above argument as a strong indication of integrability of the
spherical brane model.

In fact, the coefficients in\ \eqref{PTKJ}\ for $P_4$ and $P_6$
can be determined explicitly
from the commutativity condition\ \eqref{comm}. While ${\Bbb I}_1$
automatically commutes with any ${\Bbb I}[P]$ 
of the form \eqref{ip}, the
condition
$[\, {\Bbb I}_3\, ,\, {\Bbb I}_5\, ]
= 0$ turns out to be rather rigid. As it turns, it has
only one solution (for the coefficients $b, c, d$ in \eqref{PTKJ}) with
suitable properties\footnote{\label{jsyd}
In fact, there are exactly three
solutions. The other two correspond
to previously known systems of commuting integrals. One represents
the trivial free-field case, where all integrals ${\Bbb I}_{2k-1}$ are
quadratic in ${\bf X}$ (obviously, normalization assumed in\ \eqref{PTKJ}
has to be changed to accommodate this case); such integrals can be
compatible only with the free field boundary conditions (i.e. no
nonlinear constraints on ${\bf X}_B (\tau)$,
and possibly a term $\sim\oint
d\tau \,{\bf X}_{B}^2 (\tau)$ added to the action\ \eqref{acyl}).
Another solution corresponds to the ``KdV series'', in which the
currents $P_s (u)$ are composite fields built from the energy-momentum
tensor $T(u)=-\partial {\bf X}\cdot\partial{\bf X}$,
as was described in\ \cite{blz}.}, namely
\bea
&&b_3 = {{(N+2)}/3}\, ,\nonumber\\
&&b_5 = 3\, (N+4)/5 \label{bbb}\, , \\
&&c_5=7\, (N+4)/5\, , \nonumber\\
&&d_5 = (N+4)(36N + 59)/600 \nonumber\ .
\eea
At large $N$ the integrals ${\Bbb I}_3$ and ${\Bbb I}_5$ reduce 
to their
quadratic in
${\bf X}$ terms, in agreement with easily established
fact that in the
$N\to\infty$
limit the spherical brane model becomes a free theory\ \cite{Renn}.
Moreover, we
have checked that with this choice the integrals \eqref{sjssuyy}
indeed
satisfy
\eqref{ib} in the first nontrivial order 
of the $1/N$ expansion.

It looks likely that ones ${\Bbb I}_3$ is fixed through\ 
Eqs.\eqref{PTKJ},\eqref{bbb},
the higher-spin
integrals ${\Bbb I}_{2k-1}$ with $k=3, 4, 5\ldots$ 
are determined uniquely
by the
commutativity condition $[\, {\Bbb I}_s\,  ,\,  {\Bbb I}_3\, ]
=0$. Although for higher spins
the brute-force computation of the commutators becomes difficult, it
is possible to determine this way the coefficients
explicitly written in\ \eqref{sjssuyy}:
\bea\label{bsyast}
&&b_{2k-1} =  {{k(k-1)(2k+N-2)}\over{2(2k-1)}}\ ,  \\
&& c_{2k-1} =
{{k(k-1)(k-2)(2k+1)(2k+N-2)}\over{6\, (2k-1)}}\ . \nonumber
\eea

\subsection{Boundary state}

In view of the structure \eqref{space} of the space ${\cal H}$, it is
convenient to think of the boundary state in terms of associated
{\it boundary operator}. Natural isomorphism between ${\cal F}_{\bf P}$
and ${\bar{\cal F}}_{\bf P}$ (the right movers are replaced by the left
movers)
makes
it possible to to establish one to one correspondence between the
states in ${\cal F}_{\bf P}\otimes{\bar {\cal F}}_{\bf P}$
and operators in ${\cal F}_{\bf P}$. Thus the
boundary state $|\, B\, \rangle$ can be re-interpreted as an operator
${\Bbb B}: {\cal F}_{\bf P} \to {\cal F}_{\bf P}$. This idea was
extensively used in the study of conformal boundary conditions since
the original works \cite{Ishibashi,Cardy}. In this interpretation the
equation \eqref{deriv}
reduces to the statement of commutativity,
\bea\label{commut}
[\, {\Bbb B}\, ,\, {\Bbb  I}_s\, ] = 0\,.
\eea
The boundary state therefore can be written as
\bea
|\,  B\,  \rangle = \int_{\bf P}\, \sum_n \, B_n ({\bf P}) \ |\, n\, ,\,
{\bf P}\,  \rangle \otimes \overline{|\, n\, ,\,  {\bf P}\, \rangle}\,,
\eea
where $|\, n\, ,\,  {\bf P}\, \rangle$ are the (orthonormalized)
simultaneous
eigenvectors of the operators ${\Bbb I}_s$ in the space 
${\cal F}_{\bf P}$,
and $B_n ({\bf P})$ are corresponding eigenvalues of ${\Bbb B}$,
\bea
{\Bbb B}\ |\, n\, ,\,  {\bf P}\, \rangle = B_n ({\bf P})\
|\, n\, ,\,  {\bf
P}\, \rangle\,.
\eea
The eigenvalue $B_0 ({\bf P})$ corresponding to the Fock vacuum $
|\, 0\, ,\,  {\bf P}\,  \rangle$ in ${\cal F}_{\bf P}$ coincides 
with the
overlap
$\langle\, {\bf  0}\, |\,   {\bf P}\,  \rangle$.
This structure of $|\,  B\,  \rangle$
emphasizes
importance of the problem of simultaneous diagonalization of the
integrals of motion ${\Bbb I}_s$. Similar problem was addressed in\
\cite{blz} for
somewhat simpler model of integrable boundary interaction -- the
minimal CFT perturbed by the boundary field $\Phi_{1,3}$. In that case
associated system of integrals ${\Bbb I}_s$ was the KdV series (see
footnote $\#$\ref{jsyd}). 
It was observed in \cite{blz} that the eigenvalues of
${\Bbb I}_s$ are
related to corresponding eigenvalues of ${\Bbb B}$ in
the following way.
When the length of the boundary $2\pi R \to
\infty$, the operator $\log {\Bbb B}$ admits asymptotic expansion
in terms of the local
integrals ${\Bbb I}_s$\footnote{
This structure appears 
in integrable boundary flows down to the ``basic''
conformal
boundary, the one  which
admits  no primary boundary fields but the
identity.
In those cases 
the expansion in ${\Bbb I}_s$ corresponds to expansion of the infrared
effective
action in terms of descendents of the identity. Clearly, the Dirichlet
boundary ${\bf X}_B =0$ is of that kind. For integrable boundary flows
down to
``elevated'' fixed points, which admit nontrivial boundary primaries,
the asymptotic expansions of ${\Bbb B}$ at large $R$ should  contain
``dual'' nonlocal integrals of motion \cite{blzzz} as well. 
This is what we would expect
to have in the case of $\theta = \pi$ of the circular brane.}.
We expect similar
relation to hold in the case of the spherical brane model, namely
\bea\label{Bexp}
{\Bbb B}\, \simeq\, {\Bbb B}_{IR}\ \exp\bigg\{- \sum_{k=0}^{\infty}\,C_k \
(2\pi E^{*})^{1-2k}\ \ {\Bbb I}_{2k-1}\,
\bigg\}\, ,
\eea
where ${\Bbb B}_{IR}$ is the boundary  operator associated with the
infrared fixed point,
$E^{*}$ is the energy scale\ \eqref{succept},
${\Bbb I}_{2k-1}$ with $k\geq 1$ are
the integrals\ \eqref{PTKJ},\eqref{sjssuyy},
${\Bbb I}_{-1} \equiv \oint {\textstyle{{\rm d}\tau\over  2\pi}} = R$, and
the coefficients $C_{k}$ are just numbers, independent of the
parameters $g_0$, $R$, and ${\bf P}$.
The expansion
\eqref{Bexp} is expected to 
hold in the sense of asymptotic $R\to \infty$
series (note that ${\Bbb I}_{2k-1} \sim R^{1-2k}$ by dimensional
counting). The dependence of 
the matrix elements\ \eqref{Bexp} on $R$ and
${\bf P}$ comes through the integrals ${\Bbb I}_{2k-1}$. 
For instance, for
the vacuum-vacuum matrix element
\bea
I_{2k-1}\equiv
\langle\, 0\, ,\, {\bf P}\,  |\,{\Bbb I}_{2k-1}\, |\,0\, ,\,
  {\bf P}\,  \rangle
\eea
we have, from
Eqs.(\ref{PTKJ}-\ref{bsyast})
\bea\label{imsv}
&&I_1= {1\over R}\,\Big(\, h^2 + {N\over 24}\, \Big)\,,\nonumber\\
&&I_3
= {1\over R^3}\,\Big(\, h^4 + {{N+2}\over {12}}\,h^2 +
{{N(N+2)}\over 320}\, \Big)\,,\\
&&I_5= {1\over R^5}\,\Big(\, h^6 + {{N+4}\over 8}\,h^4+
{{(N+4)(37N+78)}\over 4800}\,h^2\nonumber\\ &&\ \ \ \ \ \
\ \ \ \ \ \ \ \ +{N(N+4)(143N+282)\over 483840}\, \Big)\,,\nonumber
\eea
and for general $k$
\bea\label{sbbsgt}
I_{2k-1} = {1\over{R^{2k-1}}}\Big(\, h^{2k} + {{k(2k+N-2)}\over
24}h^{2k-2}+\ldots\, \Big)\, ,
\eea
where $h^2=-{\textstyle{{\bf P}^2\over 4}}$. In the limit when $R\to\infty$ with
$H={\textstyle{h\over 2\pi R}}$ kept fixed, these eigenvalues become simply
$I_{2k-1}\to R\,(2\pi H)^{2k}$, and therefore  $C_k$ in
\eqref{Bexp} are just the coefficients of the power series 
expansion of
the boundary energy ${\cal E}(H)$ in\ \eqref{sbdyt},
\bea\label{frexp}
{\cal E}
(H) =E^*\  \sum_{k=0}^{\infty}\,C_k \ \big({\textstyle{ H\over  E^*}}\big)^{2k}\, .
\eea
At the moment we have only a conjecture about exact form of these
coefficients in the spherical brane model,
\bea\label{snydyy}
C_k = (-1)^k\ \ (2\Delta)^{1-k}\ \ 
{\Gamma((2k-1)\Delta)\over k!\   \Gamma^{2k-1}(\Delta)}\
\ (2k-1)^{k(1-
2\Delta)+\Delta-1}\ ,
\eea
with $\Delta={1\over N-1}$ and $k\geq 1$.
It is possible to check that 
this expression agrees with the leading order
of $1/N$ expansion in the $O(N)$ spherical brane model. The
conjecture\ \eqref{snydyy} concerns
with the coefficients $C_k$ with $k\geq 1$, but
it does not directly apply to the coefficient $C_0$.
According to\ \eqref{frexp},
the
latter relates to the boundary energy at zero external field $H$,
${\cal E}(0) = C_0\,E^*$. Presently, we do not have definite idea
what
exact value of $C_0$ is. Simply setting $k=0$ in\ \eqref{snydyy}\ is
not quite
acceptable,
since doing so results in complex numbers for generic $N$.
On the other hand, it
is likely that exact $C_0$ is closely related to the expression
\ \eqref{snydyy} with
$k=0$ (its real part $=2\pi\cot(\pi\Delta)$?). One indication is that
\eqref{snydyy} with $k=0$ develops a
pole at $N=2$, presumably signifying the small-instanton divergence
specific
to this value of $N$. Similar 
pole at $N=3$ is present in the bulk vacuum
energy of $O(N)$ sigma model \cite{AlZ}.

\end{document}